\documentstyle[12pt]{article}
\newcommand{\beq}{\begin{equation}}
\def\leta{\lambda_{\eta}}
\def\meta{m_{\eta}}
\def\mh{m_H}
\def\glq{g_{lq}}
\def\mlq{m_{lq}}
\def\gt{g_t}
\def\eonemu{\epsilon_{1\mu}}
\def\etwonu{\epsilon_{2\nu}}
\def\eonenu{\epsilon_{1\nu}}
\def\etwomu{\epsilon_{2\mu}}
\def\eone{\epsilon_1}
\def\etwo{\epsilon_2}
\def\lam{\lambda}
\def\mt{m_t}
\def\gmmu{\gamma_{\mu}}
\def\gmnu{\gamma_{\nu}}
\def\gm{\gamma}

\newcommand{\eeq}{\end{equation}}
\newcommand{\bea}{\begin{eqnarray}}
\newcommand{\eea}{\end{eqnarray}}

\def\ie{\hbox{\it i.e.}{}}    
\def\eg{\hbox{\it e.g.}{}}    

\def\etal{\hbox{\it et al.}{}}
\relax

\def\figcap{\section*{Figure Captions\markboth
     {FIGURECAPTIONS}{FIGURECAPTIONS}}\list
     {Fig. \arabic{enumi}:\hfill}{\settowidth\labelwidth{Fig. 999:}
     \leftmargin\labelwidth
     \advance\leftmargin\labelsep\usecounter{enumi}}}
 \relax
\def\reflist{\section*{REFERENCES\markboth
     {REFLIST}{REFLIST}}\list
     {[\arabic{enumi}]\hfill}{\settowidth\labelwidth{[999]}
     \leftmargin\labelwidth
     \advance\leftmargin\labelsep\usecounter{enumi}}}
 \relax
\def\tabcap{\section*{Tables\markboth
     {TABLES}{TABLES}}\list
     {Table \arabic{enumi}:\hfill}{\settowidth\labelwidth{Table 999:}
     \leftmargin\labelwidth
     \advance\leftmargin\labelsep\usecounter{enumi}}}
 \relax

\begin{document}
\begin{titlepage}
 \null
 \vskip 0.5in
\begin{center}
\makebox[\textwidth][r]{IP/BBSR/99-6}

 \vspace{.15in}
  {\Large \bf
    Leptoquark Contribution to the Higgs Boson 
                     Production at the LHC Collider}
  \par
 \vskip 1.5em
 {\large
  \begin{tabular}[t]{c}
    Pankaj Agrawal  \\
\em Institute of Physics \\
\em Sachivalaya Marg \\
 \em Bhubaneswar, Orissa 751005 India\\
   \em and\\
 Uma Mahanta \\
 \em Mehta Research Institute \\
  \em Chhatnag Road, Jhusi \\
  \em Allahabad, UP 211019 India\\
   \\
  \end{tabular}}
 \par \vskip 5.0em

 {\large\bf Abstract}
\end{center}
\quotation
In this report we study  how  a light-scalar leptoquark could
affect the Higgs
boson production cross-section at the LHC collider. We construct the most
 general renormalizable
and gauge invariant effective Lagrangian involving the standard model
particles and a scalar, isoscalar leptoquark, $\eta$.
 The total cross-section for $pp\rightarrow H+X$ is then calculated
for different values of the unknown parameters $\leta$, $\meta$ and
$\mh$.(Here $\leta$ is the coupling associated with the 
Higgs-leptoquark interaction.)

We find that if $\leta$ is moderately large and $\meta$ is around a few
hundred GeV, then the cross-section is significantly larger than
the standard model value.

\endquotation
\baselineskip 10truept plus 0.2truept minus 0.2truept 
\vfill 
\mbox{Feb 1999} 
\end{titlepage} 
\baselineskip=21truept plus 0.2truept minus 0.2truept 
\pagestyle{plain}
\pagenumbering{arabic}

Leptoquarks (LQ) occur naturally
in many extensions of the standard model, \eg, 
grand unified models, technicolor models 
and composite models of quarks and leptons. However
 if leptoquarks couple to the quark pair as well as to the quark-lepton 
pair then they give rise to rapid proton decay.
This in turn requires that the leptoquark should be heavy, $m_{lq}\approx 
10^{12}-10^{15}$ GeV [\ref{lq1}].
To evade this catastrophic strong bound it is usually 
assumed that leptoquarks couple only to the quark-lepton pair. To enable them
to couple to a quark-lepton pair, leptoquarks must form a color triplet or
antitriplet. They can form a singlet or doublet or triplet representation
of weak SU(2) group and also carry weak hypercharge. The couplings and masses
of leptoquarks that couple to quark-lepton pair are constrained by a variety
of low energy processes, \eg, pion and kaon decays [\ref{lq1}],
 B and D meson decays,
precision electroweak measurements at Z boson pole [\ref{lq2}],
 Tevatron  and HERA data. The Tevatron sets limits [\ref{lq3},
 \ref{lq3a}] on $m_{lq}$
(the mass of the leptoquark)
from pair production processes. These bounds depend on the color and 
electroweak (EW) quantum numbers of the LQ. The bounds are most stringent 
($>$ 225 GeV) for the first generation LQ and become progressively
weaker for the second ($>$ 131 GeV) and third generation ($>$ 95 GeV) LQs. 
HERA sets limits [\ref{lq3}, \ref{lq3b}] on $\mlq$ from single production of
leptoquarks. These limits depend on the value of $\glq$, the Yukawa like 
coupling of the LQ to quark-lepton pair.
 For $\glq =
e$, the limits on a scalar, weak isoscalar, Q=-${1\over 3}$, LQ
are $\mlq >$ 237 GeV for the first generation and $\mlq >$ 73 GeV for the
second generation. In this article we shall consider a third generation
leptoquark whose coupling to q-l pairs of the first two generations
are very small. This assumption will enable us to consider
leptoquarks in the 100-200 GeV mass range and still be consistent 
with the above bounds. However, we note that if some
crucial assumptions that go into arriving at the Tevatron bounds,
\eg, $B(eq) = 1$, can be relaxed, then second/first generation
leptoquarks could still exist in $100 - 200$ GeV range.
 For example mixing between 
different multiplets of LQ's could reduce the branching ratios
considerably.
Furthermore since

the coupling $\glq$ will not enter our computations our results
will also be valid for just a color triplet scalar. 

As shown later the most general effective Lagrangian involving the
standard model particles and the LQ includes an interaction term 
between the LQ
and the Higgs boson. In the presence of this interaction 
the coupling $\glq$ could contribute to $pp\rightarrow H + X$
 through a loop diagram made up of a lepton and a leptoquark
with the Higgs boson attached to the leptoquark line. The incoming quark and
antiquark states are to be attached at the lepton and leptoquark junctions.
$\glq$ could also contribute to $pp \rightarrow l^{+}l^{-} + H + X $ through  
a tree level diagram.
However it is not hard to see that the 
coupling $g_{lq}$ of second or third generation LQ
to quark-lepton pairs of first generation 
cannot play a significant role in 
Higgs boson production in our model at a pp collider.
As an example, let us consider a third generation leptoquark.
Then, first the coupling $\glq$ of third generation LQ to q-l
pairs of first generation is expected to be  very small ($\glq\ll e$).
 Second a smaller luminosity of  $q \bar q$ 
configurations at the LHC makes the contribution  of $\glq$ to 
the Higgs boson production at the LHC
much smaller than those computed here.

At the LHC, the dominant production mechanism for the standard model
Higgs boson is through gluon-fusion mechanism [\ref{hig1}] (see
Fig. 1). If light LQs
exist, there will be additional diagrams with the leptoquark 
loops (see Fig. 2).
In this brief report, we shall compute the effect of these additional 
diagrams on the the Higgs boson production via this mechanism. 
In order to be consistent with the HERA bounds on $\mlq$
we shall assume that $\glq$ is much smaller than e
 so that $\mlq$ could
lie in the range $100 - 200$ GeV. We shall then show that a low
 energy effective 
interaction between the higgs doublet and a scalar leptoquark can 
significantly modify the Higgs boson production cross-section at the LHC 
provided the corresponding coupling $\leta$ (the coupling 
associated with the Higgs-leptoquark interaction) is large enough, 
\ie, of the same 
order as the yukawa coupling of the top quark, $\gt$. 
The phenomenological advantage of using the 
coupling $\leta$ is that unlike $\glq$
 it is not unduly constrained by low energy 
experiments. For simplicity let us consider a scalar, weak isoscalar, 
$Q=-{1\over 3}$ leptoquark $\eta$.
The form of the effective interaction between the LQ and the higgs scalar
will depend on the EW quantum numbers of $\eta$ and therefore they 
do affect the cross-section for the Higgs boson production. 

If $\mlq$ lies in the few hundred GeV range then the low energy 
effective Lagrangian suitable for describing physics in this energy range
should involve $\eta$ besides the usual SM particles. We shall 
require the low
energy lagrangian to be renormalizable and invariant under $SU(3)_c\times
SU(2)_l\times U(1)_y$. We then have

\begin{eqnarray}
  \label{eq:1a}
L_{eff}= \tilde{L}_{sm} +(D_{\mu}\eta )^+(D^{\mu}\eta )-V(\phi, \eta )
+L_{ql}.
\end{eqnarray}

 Here $L_{sm} = {\tilde L}_{sm}-V(\phi )$, $V(\phi )= -\mu^2 \phi^+\phi
+\lambda (\phi^+\phi )^2$,

\begin{eqnarray}
  \label{eq:1b}
V(\phi, \eta ) = V(\phi )+ m^2_{0\eta }\eta^+\eta-
\leta (\phi^+\phi )\eta^+\eta +{\lambda^{\prime}\over 4}
(\eta^+\eta )^2,
\end{eqnarray}

and,

\begin{eqnarray}
  \label{eq:1c}
L_{ql} =[g_l{\bar l}_L i\tau_2 q^c_L+g_r{\bar e}_R u^c_R ]\eta + h.c.
\end{eqnarray}

Here $\leta$ is the dimensionless coupling associated with the 
Higgs-leptoquark interaction term.
The LQ mass gets shifted from $m_{0\eta }$ to $m_{\eta }$ after the
electroweak symmetry 
breaking where $m^2_{\eta }= m^2_{0\eta }-\leta {v^2\over 2}$. The Lagrangian
 $L_{ql}$ gives the coupling of the third generation LQ  $\eta$ 
 to the q-l pair
of the third generation. Its coupling to quark-lepton pairs of first
two generations has been neglected.
In the unitary gauge the interaction term between $\phi$  and $\eta$ can
be written  as $\leta (\phi^+\phi )(\eta^+\eta )=
{\leta \over 2} (v^2 +2vh +h^2)\eta^+\eta$.
 Note that the above effective Lagrangian
includes all possible terms subject to the constraints of
renormalizability and gauge invariance and the  assumption that $\eta$
is an isoscalar.
If $\eta$ were a weak isodoublet then we should add the term 
$\lam ^{\prime}_{\eta}(\phi^+\eta )(\eta^+\phi )$ to the effective
Lagrangian. In unitary gauge this term can be expressed as
$\lam^{\prime}_{\eta}{(v+h)^2\over 2}\eta^{+a2}\eta^{a2}$ where a
is the color index and 2 stands for the $I_3=-{1\over 2}$ component
of $\eta$. After the electroweak symmetry breaking the term proportional to
$\lam^{\prime}_{\eta}$ causes the h boson to interact only with the down component
of the LQ. Whereas the term proportional to $\lam_{\eta}$ causes
the h boson to interact with both the isospin components of $\eta$
with equal strength.
The scalar potential of the standard model $V(\phi )$ is bounded from below for
$\lambda >0$. Further it produces a vacuum with the desired symmetry
breaking properties. Here we need to find the conditions to be satisfied by
the parameters of $V(\phi, \eta )$ so that it exhibits the same properties.
In order that $V(\phi, \eta )$ is bounded from below the quartic part of
it must be positive. It can be shown that this can happen if 
$\lambda >0$, $\lambda^{\prime}>0$ and $\leta <0$ or
$\lambda >0$, $\lambda^{\prime}>0$ and $0<\leta <2\sqrt 
{\lambda\lambda^{\prime}}$. Finally in order that the vacuum does not
break the $SU(3)_c$ symmetry $m_{0\eta }$ must satisfy the condition
$m^2_{0\eta }-\leta {v^2\over 2}>0$.

In the context of the standard model, the diagrams contributing to the
Higgs boson production at the LHC via gluon fusion mechanism are displayed
in Fig. 1. These diagrams involve a top quark loop with two
gluons and a Higgs boson attached to it at the interaction vertices. The 
contribution of the diagrams to the amplitude is given by 
$M= M_1+M_2$ where

\begin{eqnarray}
M_1 & = & -{i\gt g_s^2\over 2{\sqrt 2}}\eonemu (q_1, \lam_1)
\etwonu (q_2, \lam_2)
\delta_{ab}\int {d^4l\over (2\pi)^4} {N_1\over D_1}, \nonumber \\
M_2 & = & -{i\gt g_s^2\over 2{\sqrt 2}}\eonenu (q_1, \lam_1)
\etwomu (q_2, \lam_2)
\delta_{ab}\int {d^4l\over (2\pi)^4} {N_2\over D_2}, \nonumber \\
{N_1\over D_1}& = &{(l.\gm+\mt)\gmmu ((l+q_1).\gm+\mt)((l-q_2).\gm+\mt)\gmnu
\over (l^2-\mt^2)[(l+q_1)^2-\mt^2][(l-q_2)^2-\mt^2)]}, \nonumber \\
{N_2\over D_2}& = &{(l.\gm+\mt)\gmmu ((l+q_2).\gm+\mt)((l-q_1).\gm+\mt)\gmnu
\over (l^2-\mt^2)[(l+q_2)^2-\mt^2][(l-q_1)^2-\mt^2)]}.  \nonumber
\end{eqnarray}

    In the notations of Ref 5, we can write the total amplitude as,
\begin{eqnarray}
  \label{eq:2} 
     M & = & {\gt \over 2{\sqrt 2}}{g_s^2 \over 16 \pi^2}\eonemu (q_1, \lam_1)
\etwonu (q_2, \lam_2) \delta_{ab}
   [32 \mt C_{\mu \nu}(q_1, q_2, \mt, \mt, \mt) \eone^\mu \etwo^\nu  \nonumber \\
    & & + 16 \mt ( q_1 . \etwo \eone^\mu  C_{\mu}(q_1, q_2, \mt, \mt, \mt) 
    + q_2 . \eone \etwo^\mu C_{\mu}(q_2, q_1, \mt, \mt, \mt) )
    \nonumber \\
 & & +  4 \mt (-  q_1 . q_2 \eone . \etwo + q_1 . \etwo q_2 . \eone  )
      (C_{0}(q_1, q_2, \mt, \mt, \mt)  + C_{0}(q_2, q_1, \mt, \mt,
      \mt))
              \nonumber \\
 & &  -  8 \eone . \etwo \mt B_{0}(q_1 + q_2, \mt, \mt)].
\end{eqnarray}
In the above $\epsilon_{1\mu}(q_1, \lambda_1)$ and 
$\epsilon_{2\nu}(q_2, \lambda_2)$ are the  polarization vectors of the
incoming gluons. $\lambda_1$ and $\lambda_2$ stand for the polarization 
states of the gluons and $l.\gamma = l^{\mu}\gamma_{\mu}$. Here $B_{0},
C_{0}, C_{\mu},$ and $C_{\mu \nu}$ are scalar and tensor integrals, as
defined in Ref 5. As noted below, we use the techniques of [\ref{ov}]
to reduce the tensor integrals $C_{\mu}$ and $C_{\mu \nu}$ to scalar 
integrals.

Next consider the contribution arising from LQs. The coupling $\glq$
could contribute to the process $pp\rightarrow H+X$ through the parton level
subprocess $q{\bar q}\rightarrow H$ that involves a loop diagram made
up of a lepton and a LQ. However since for third generation LQ 
at a pp collider the relevant $\glq\ll e$  
the contribution of this subprocess can be neglected in comparison
to that of $gg\rightarrow H$.
Leptoquarks have two distinct types of interaction vertices with gluons.
In the first kind a single gluon line meets two lines. It is a derivative 
coupling 
and the Feynman rule for it is $-ig_s(p_{1\mu}+p_{2\mu})T_a$ where
$p_1$ and $p_2$ are the incoming and outgoing momenta along the $\eta$ lines.
In the second kind of vertex two gluon lines meet two $\eta$ lines.
It is a dimension four coupling and the Feynman rule for it is
$ig_s^2(T_aT_b+T_bT_a)g_{\mu\nu}$. Since leptoquarks form a fundamental 
representation of color, the same color Gellmann matrices appear in the
above Feynman rules.  Finally the Feynman rule for the interaction 
vertex of the Higgs boson with two $\eta$ lines is $i\leta v$.
 These rules give rise to three distinct Feynman diagrams for 
leptoquark contribution to the Higgs boson production. The 
corresponding  amplitudes are given by

\begin{eqnarray}
 M_1^{\prime}&  = & {i\over 2}g_s^2\leta v\delta_{ab}\int {d^4l\over (2\pi)^4}
{(2l+q_1).\eone (2l+2q_1 +q_2).\etwo\over
(l^2-\meta^2)[(l+q_1)^2-\meta^2][(l+q_1+q_2)^2-\meta^2]}, \nonumber \\
M_2^{\prime}& = &{i\over 2}g_s^2\leta v\delta_{ab}\int{d^4l\over (2\pi)^4}
{(2l+q_2).\etwo (2l+2q_2
+q_1).\eone\over
(l^2-\meta^2)[(l+q_2)^2-\meta^2][(l+q_1+q_2)^2-\meta^2]},
             \nonumber \\
M_3^{\prime}& = & -ig_s^2\leta v\delta_{ab}\int {d^4l\over (2\pi)^4}
{\eone.\etwo\over (l^2-\meta^2)[(l+q_1+q_2)^2-\meta^2]}. 
\end{eqnarray}

  As earlier, in the notations of the Ref. 5, 
  \begin{eqnarray}
    \label{eq:3}
 M^{\prime} & = & M_1^{\prime} +  M_2^{\prime} + M_3^{\prime} \nonumber \\
      &  = & - {1 \over 2} {g_s^2 \over 16 \pi^2}  \leta  v\delta_{ab}
          [4 \eone^\mu \etwo^\nu ( C_{\mu \nu}(q_1, q_2, \meta, \meta,
          \meta) + C_{\mu \nu}(q_2, q_1, \meta, \meta, \meta) )
          \nonumber \\
  & &       + 4 \eone^\mu q_1 . \etwo  C_{\mu}(q_1, q_2, \meta,
                     \meta, \meta) 
            + 4 \etwo^\mu q_2 . \eone    C_{\mu}(q_2, q_1, \meta,
                     \meta, \meta)       \nonumber \\
 & &               - 2 \eone . \etwo B_{0}(q_1
                   + q_2, \meta, \meta) ]   
  \end{eqnarray}

Note that the amplitudes $M_1^{\prime}$, $M_2^{\prime}$ $M_3^{\prime}$
are separately logarithmically divergent. However one can check that the
individual logarithmic divergences cancel in the sum 
$M_1^{\prime}+M_2^{\prime}+
M_3^{\prime}$. The parton level standard model cross section can be determined
from $\vert M_1+M_2 \vert ^2$. On the other hand the parton level
total cross section 
that includes the leptoquark contribution
is to be determined from $\vert M_1+M_2+\delta M \vert ^2 $ 
where $\delta M =M_1^{\prime}+M_2^{\prime}+M_3^{\prime}$
is the leptoquark contribution. Note that the nature of interference 
between the standard model amplitude and leptoquark amplitude reverses 
with the sign of $\leta$. However in this work we shall
consider only positive values of $\leta$ which gives rise to constructive
interference with the standard model amplitude.
Further to average over  polarization and color of incoming
gluons we need to divide the parton level cross section 
by a factor of 256. After determining the parton 
level cross sections, 
we integrate them with
the gluon density functions appropriate for the LHC environment.

To compute the amplitude, we have  to calculate the loop integrals.
The method of our calculation can be found in  [\ref{penta1}].
In brief, we have used Oldenborgh-Vermaseren techniques [\ref{ov}] to reduce
the tensor integrals to scalar integrals. Here to complete the calculation
we need only two scalar integrals. We have checked the cancellation of
the ultraviolet divergences numerically as well as analytically. Afterwards,
we evaluate the amplitude and its appropriate square numerically.
 We have convoluted the parton level cross-section with the set $3$
( leading-order fits) of the CTEQ4 gluon distribution functions [\ref{cteq1}] 
which have been evolved 
to $Q^2 = {\hat s} = m^2_H$. Some results of our calculation are shown in Fig.
3 and Fig. 4. In Fig. 3 we show how the total cross section $(\sigma_t)$
and the standard model cross-section $\sigma_{sm}$ for the physical process 
$p + p \rightarrow H + X$ vary with $m_H$ for $\meta = 150$ GeV and 
$\leta =.7$. From the graph we find that at $m_H $ = 150 GeV, 
$\sigma_t\approx $ 10.5 pb whereas $\sigma_{sm}\approx $ 6.25 pb which 
implies an enhancement of nearly 70\% due to leptoquark contribution. 
In Fig. 4 we keep $m_H$ fixed at 150 GeV and show how the total cross section
varies with $\meta$ for two different values of $\leta$ namely .8 and .4.
Our results show that light leptoquarks ($\meta \approx 150-200 $ Gev )
with large enough coupling to standard model Higgs boson ($\leta\approx g_t$)
can lead to a significant
 enhancement in the Higgs boson production cross section relative to that
in the standard model. 

The major sources of uncertainties in our calculations are values of 
gluon distributions [\ref{cteq2}] and the scale at which these are evaluated. 
These overall uncertainties are expected to be of the order of $20-30\%$.
We estimate this by varying the Q and also using leading order MRST
parton distributions [\ref{mrs}]. However, this uncertainty
would not drastically affect the relative results (\ie, between purely
standard model results and the results with leptoquarks contributions).
We have presented the
 leading order results. Higher order QCD corrections are known
to modify the standard model results quite significantly. For
example the next to leading
order QCD corrections increase the cross section by around 30 \% [\ref{hig2}]
over a
large range of higgs mass.
 However, higher order
QCD corrections will modify the additional leptoquarks contribution in
a similar way. Therefore, we would not expect the higher order QCD
 corrections to significantly alter our main conclusions.

The interaction term between the Higgs boson and the LQ $\eta$ given 
by eqn.(2) could modify the Higgs boson decay properties. If
$m_H>2m_{\eta }$ then the
Higgs boson will decay into a pair of LQ's of the second or third
generation. This will modify the branching ratios of various Higgs
boson decay modes. The consequences of this modifications will depend
on the parameter $\lambda_\eta$. 
Each LQ will decay into a q-l pair of the corresponding generation. 
One useful new signature of Higgs boson production and its
consequent decay will be
``$2$ jets+ $2$ leptons'' with peaks in the mass distributions
of each $q-l$ pair. In addition to this there will
be other interesting signatures of the Higgs
boson production.

In conclusion, we have shown that light leptoquarks
of the third generation  can 
significantly modify the cross-section for the Higgs boson production 
through the process $pp\rightarrow H+X$ provided  its coupling to the Higgs
boson is strong enough ( \ie, of the order of the yukawa coupling 
of the top quark). We have shown that for $\leta =.7$ and $\meta =150 $ GeV
the cross-section increases by as much as 70 \% relative to its standard
model value. The coupling $\leta$ is not constrained by available low 
energy as well as collider data. In fact the experimental conditions at 
the LHC will offer the first
 opportunity for probing such a coupling between
 light leptoquarks and the Higgs boson, if such particles exist.
Our results will also apply to second or first generation leptoquarks
if the branching ratios assumed in arriving at the 
 Tevatron bounds are relaxed. More generally,
our results will be applicable to color triplet scalar particles that interact
with the Higgs boson in the manner prescribed in the text.

\relax
\def\pl#1#2#3{
     {\it Phys.~Lett.~}{\bf B#1} (19#3) #2}
\def\zp#1#2#3{
     {\it Zeit.~Phys.~}{\bf C#1} (19#3) #2}
\def\prl#1#2#3{
     {\it Phys.~Rev.~Lett.~}{\bf #1} (19#3) #2}
\def\rmp#1#2#3{
     {\it Rev.~Mod.~Phys.~}{\bf #1} (19#3) #2}
\def\prep#1#2#3{
     {\it Phys.~Rep.~}{\bf #1} (19#3) #2}
\def\pr#1#2#3{
     {\it Phys.~Rev.~ }{\bf D#1} (19#3) #2}
\def\epj#1#2#3{
     {\it Euro.~Phys.~Jour. }{\bf C#1} (19#3) #2}
\def\np#1#2#3{
     {\it Nucl.~Phys.~}{\bf B#1} (19#3) #2}
\def\ib#1#2#3{
     {\it ibid.~}{\bf #1} (19#3) #2}
\def\nat#1#2#3{
     {\it Nature (London) }{\bf #1} (19#3) #2}
\def\ap#1#2#3{
     {\it Ann.~Phys.~(NY) }{\bf #1} (19#3) #2}
\def\sj#1#2#3{
     {\it Sov.~J.~Nucl.~Phys.~}{\bf #1} (19#3) #2}
\def\ar#1#2#3{
     {\it Ann.~Rev.~Nucl.~Part.~Sci.~}{\bf #1} (19#3) #2}
\def\ijmp#1#2#3{
     {\it Int.~J.~Mod.~Phys.~}{\bf #1} (19#3) #2}
\def\cpc#1#2#3{
     {\it Computer Physics Commun. }{\bf #1} (19#3) #2}
\begin{reflist}

\item \label{lq1} W. Buchmuller, Acta Phys. Austriaca, Suppl 27, 517 (1985);
 W. Buchmuller and D. Wyler, \pl{177}{377}{86};
W. Buchmuller, R. Ruckl and D. Wyler \pl{191}{442}{87}.

\item \label{lq2} P. Langacker, M. Luo and A. K. Mann, \rmp{64}{87}
{92}; O. J. P. Eboli, M. C. Goncalez Garcia and J. K. Mizukoshi,
\np{443}{20}{95}.

\item \label{lq3} Review of Particle Properties, \epj{3}{1}{98}.

\item \label{lq3a} B. Abbott \etal (D0 Collaboration),
  \prl{80}{2051}{98}; F. Abe \etal (CDF Collaboration), \prl{78}{2906}{97};
  \ib{75}{1012}{95}.

\item \label{lq3b} S. Aid \etal (H1 Collaboration), \pl{369}{173}{96};
   M. Derrick \etal (ZEUS Collaboration), \pl{306}{173}{93}.

\item \label{hig1} H. Georgi, S. Glashow, M. Machaceck and D. Nanopoulos, 
\prl{40}{692}{78}; T. Rizzo, \pr{22}{178}{80}. S. Dawson,
Perspectives on Higgs Physics'', World Scientific (1993).

\item\label{pv} G. Passarino and M. Veltman, \np{160}{151}{79}.

\item \label{penta1} P. Agrawal and G. Ladinsky, preprint IP/BBSR/99-11.

\item \label{ov} G. J. van Oldenborgh and J. A. M. Vermaseren, Z. Phys C46,
  425 (1990).

\item \label{cteq1} H. L. Lai {\it et. al.}, \pr{55}{1280}{97}.

\item \label{cteq2} J. Huston {\it et. al.},
hep-ph/9801444 , Fermilab-Pub-98/046-T.

\item \label{mrs} A.D. Martin, R.G. Roberts, W.J. Stirling and R.S Thorne, 
           Univ. Durham preprint DTP/98/10 (1998), [hep-ph/9803445]. 

\item \label{hig2} A. Djouadi, M. Spira and P. M. Zerwas, \pl{264}{440}{91}.

\end{reflist}

\newpage

\begin{figcap}

\item Diagrams for the process $ g g \to H$ in the standard model.

\item Additional diagrams for the process $ g g \to H$ with leptoquarks
      in the model.

\item Cross-section for $p p \to H X$ with $\meta = 150$ GeV and $\leta
      = 0.7$.

\item Cross-section for $p p \to H X$ with $\mh = 150$ GeV and $\leta
      = 0.4$ and $0.8$. 

\end{figcap}


%



%
%
%


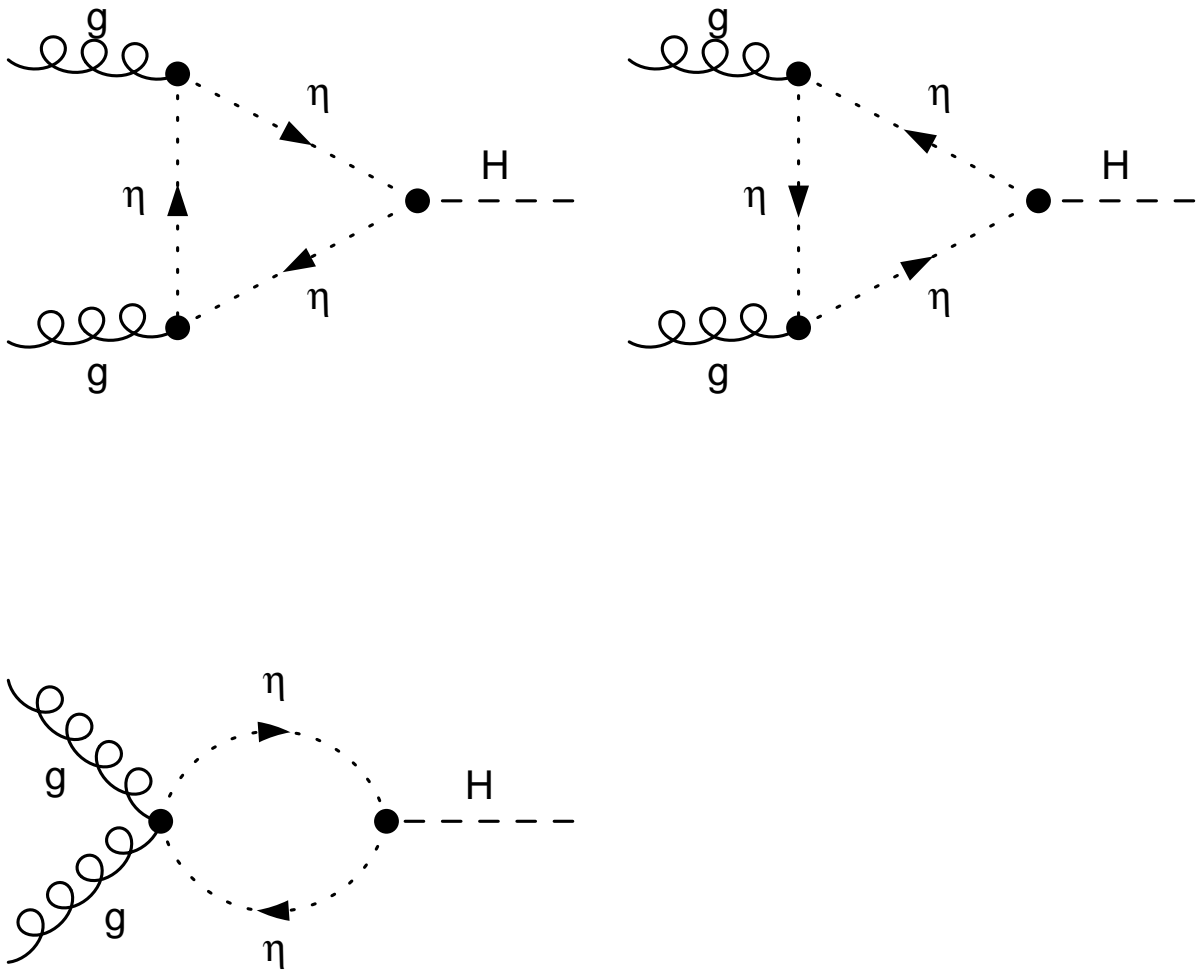
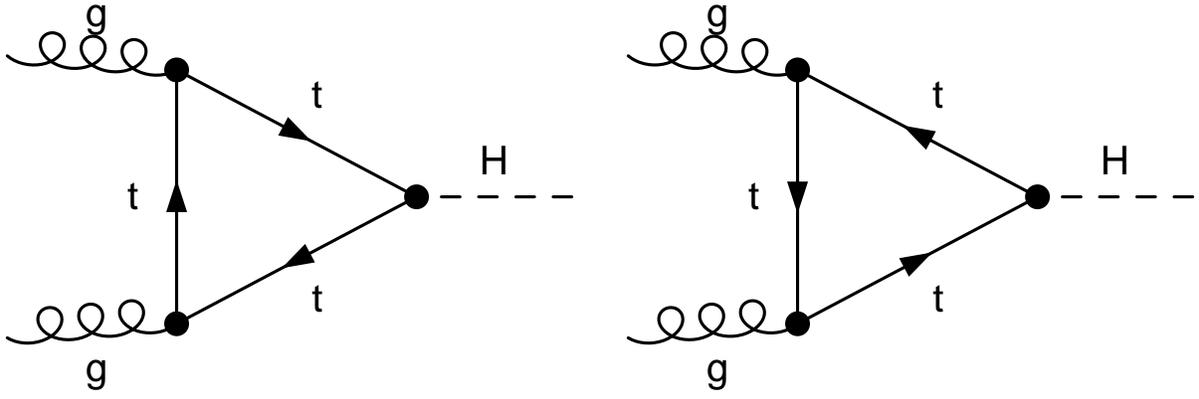
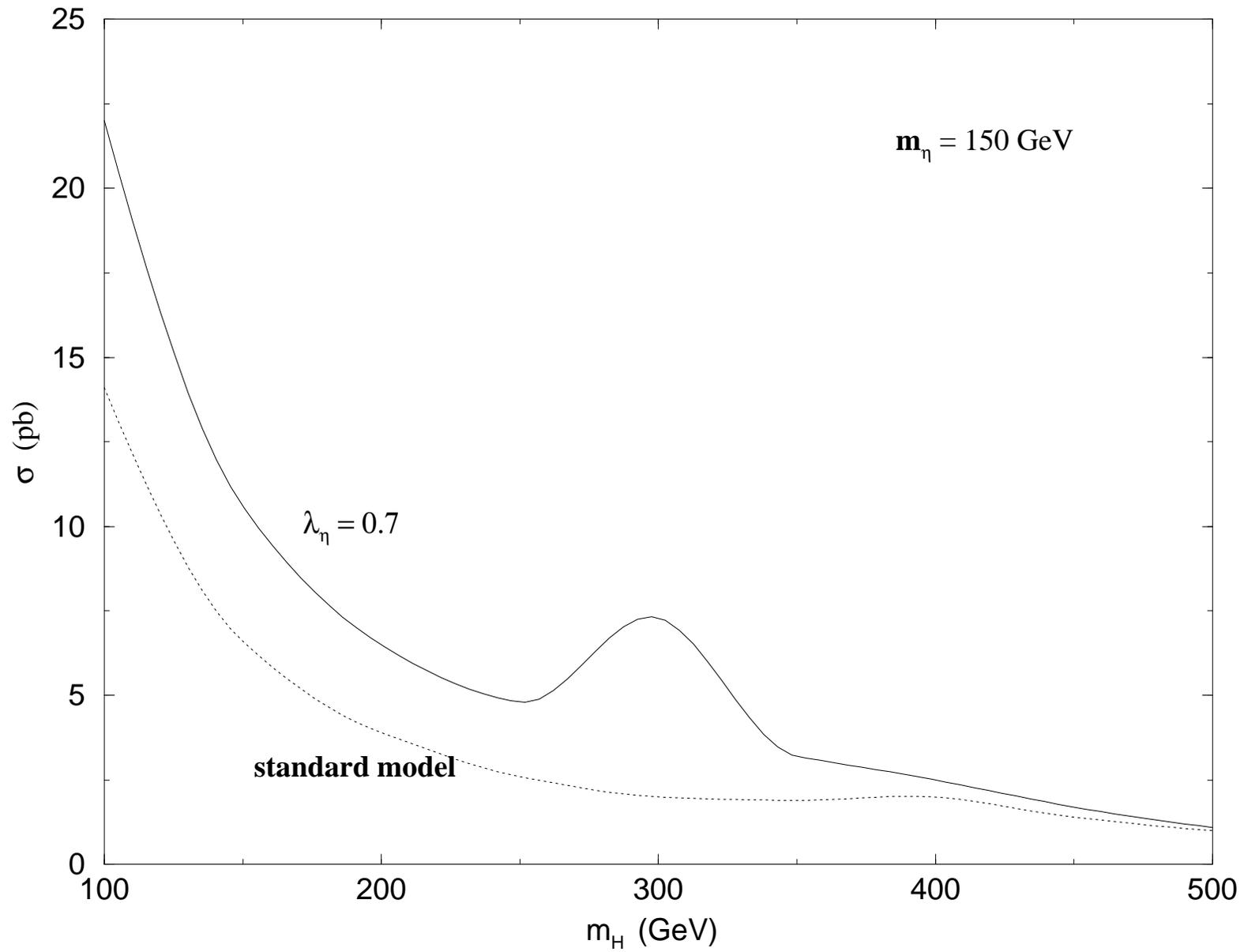
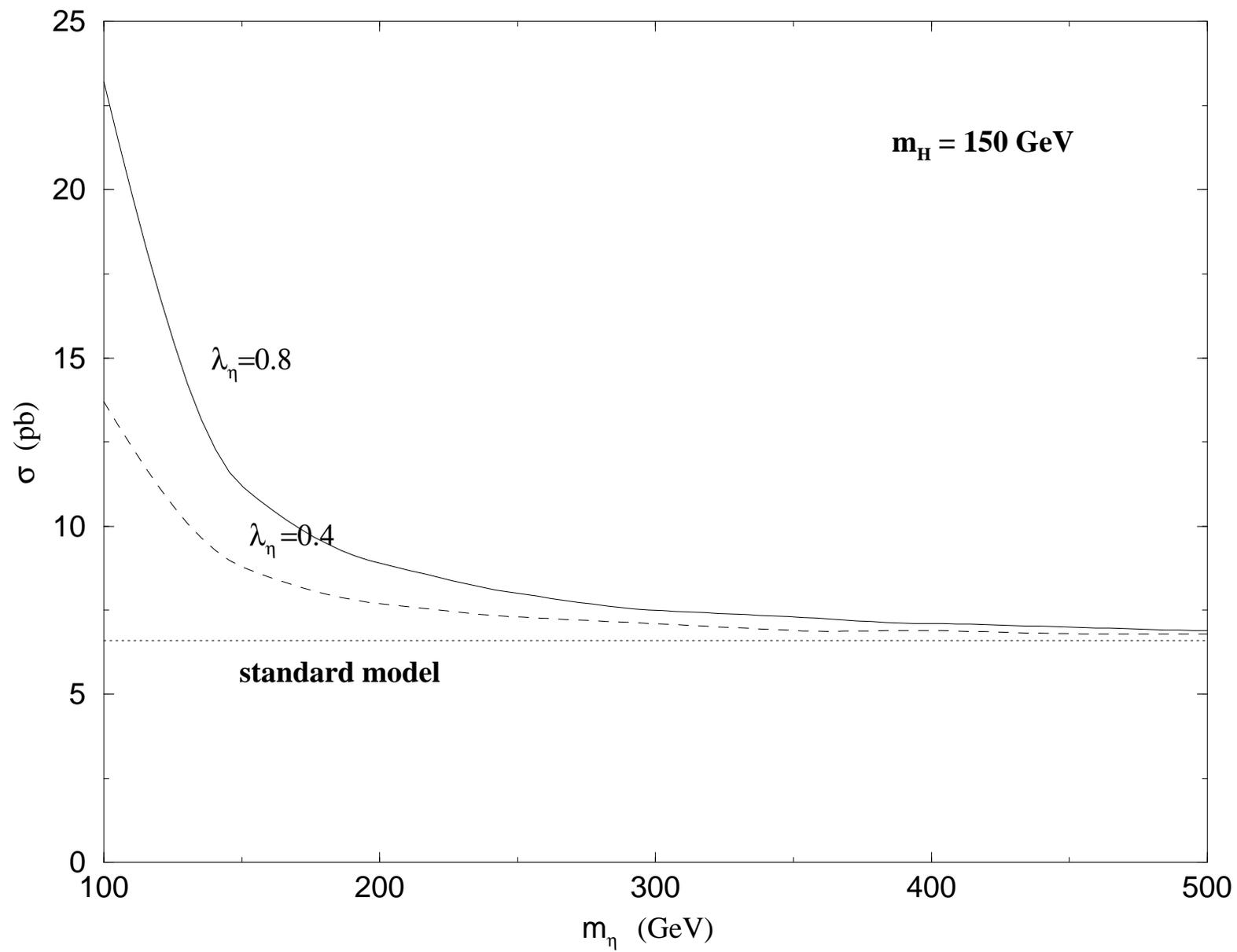
\end{document}